\documentclass{amsart}
\usepackage{graphicx}
\usepackage{amsmath}
\usepackage{color}
\newtheorem{thm}[equation]{Theorem}
\newtheorem{pro}[equation]{Proposition}
\newtheorem{cor}[equation]{Corollary}
\newtheorem{lem}[equation]{Lemma}

\theoremstyle{definition}

\newtheorem{DEF}[equation]{Definition}

\usepackage{amsfonts,amssymb,amscd,amsmath,enumerate,verbatim,calc}
\def\aa{\mathcal A}

\def\e{\epsilon}

\def\bbbf{\mathbb{F}}

\def\LL{\mathcal{L}}

\def\lam{\lambda}
\def\Lam{\Lambda}

\def\hh{\mathcal H}

\def\u{{\mathcal U}}

\def\bbbz{{\mathbb Z}}

\def\bbbk{{\mathbb K}}

\def\lm{\lambda}
\def\e{E}

\def\pp{{\mathcal P}}
\def\vv{{\mathcal V}}
\def\uu{{\mathcal U}}
\def\aa{{\mathcal A}}

\def\lam{\Lambda_I^{\bar{i}}}

\def\PP{\mathcal{P}}

\def\e{\epsilon}

\def\lam{\lambda}
\def\Lam{\Lambda}
\begin{document}

\markboth{On the structure of graded  Poisson color algebras} {V. Khalili}

\date{}

\centerline{\bf On the structure of graded  Poisson color algebras}

\vspace{.5cm}\centerline{Valiollah Khalili\footnote[1]{Department
of mathematics, Faculty of sciences, Arak University, Arak 385156-8-8349, Po.Box: 879, Iran. 
 V-Khalili@araku.ac.ir\\
\hphantom{ppp} MSC 2020: 17B63, 17B75, 17B05, 17A60
\\
\hphantom{ppp}Keywords: Poisson color algebras, graded algebras and structure theory }\;\;}

\vspace{1cm} \noindent ABSTRACT:

 In this paper we introduce the class of graded Poisson color  algebras
 as the natural generalization of graded Poisson algebras and graded Poisson superalgebras.  For $\Lam$ an arbitrary abelian  group, we show that any of such
$\Lam$-graed  Poisson color  algebra $\pp,$ with a symmetric $\Lam$-support is of the form $\pp=\uu\oplus\sum_{j}I_{j},$  with $\uu$  a subspace of $\pp_1$  and any  $I_j$ a well described graded ideal of $\pp,$  satisfying  $\{I_j, I_k\}+I_j I_k=0$  if
 $j\neq i.$  Furthermore, . under certain conditions, the gr-simplicity of $\pp$  is characterized and it is
shown that $\pp$ is the direct sum of the family of its graded simple ideals.

\vspace{1cm} \setcounter{section}{0}
\section{Introduction}\label{introduction}

A Poisson algebra is a Lie algebra endowed with a non-commutative associative product in such a way that the Lie and associative products are compatible via a Leibniz rule, which initially  appeared in  the work of Smeon-Denis Poisson in the theory of celestial mechanics \cite{A}. These algebras play a central role in the
study of Poisson geometry \cite{A1, L}, in deformation quantization \cite{H, K} and  in deformation of commutative associative algebras \cite{G}. Poisson  structures are also used in the
study of vertex operator algebras (see \cite{FB}). Moreover, the cohomology group, deformation, tensor product
and $\Gamma$-graded of Poisson algebras have been studied by many authors in \cite{GK, WGZ, WZC, ZL}.

 A Poisson   color algebra  has simultaneously an  associative color structure algebra and a Poisson color algebra structure, satisfying the Leibniz color identity \cite{WGZ}. This algebra naturally generalizing both Poisson algebra and Poisson superalgebra and motivated mainly by their applications in geometry and mathematical physics  (see \cite{AD, B, H, SS}).

The study of gradings on Lie algebras begins in the 1933 seminal Jordan’s work,
with the purpose of formalizing Quantum Mechanics \cite{J} .The descriptions of all possible gradings on algebras plays
an important role both in the structure theory of finite dimentional and infinite
dimentional algebras and its applications ( see \cite{BB, BSZ, BT, C4, C5, E, EK, Kh1,  PZ, V}). It is worth mentioning that the so-called techniques of connection of roots had long
been introduced by Calderon, Antonio J, on split Lie algebras with symmetric root
systems in \cite{C1}. Since then, the interest on gradings by the technique of connections of elements in the support of the
graing on different classes of algebras has been remarkable in the recent years, motivated in part by their application in physics and geometry. Recently, in \cite{BCNS, C2, C3,  CC,CS1, CS2, Kh2}, the structure of  graded Lie algebras, graded Lie superalgebras, graded Leibniz algebras,  graded Lie triple systems, graded Leibniz triple systems, graded Lie algebra of order 3 and graded 3-Leibniz algebras  have been determined
by the connections of the support of the grading.  

Our goal in this work is to study the structure of arbitrary graded Poisson color algebras (not necessarily
 finite-dimensional) introduced as the natural extension of graded Lie algebras  over an arbitrary base field $\bbbf$ by focusing on their structure. We extend
the results of split non-commutative  Piosson   algebras in \cite{C0} to the
framework of   arbitrary graed Poisson color algebras by the technique of connections of elements in the support of the
grading. 

Throughout this paper,graded Poisson color algebras $\pp$ are considered arbitrary dimension and over an
arbitrary base field  $\bbbf,$ with characteristic zero. We also
consider an additive abelian group $G$ with identity zero and an arbitrary abelian group $\Lam$ with juxtaposition product and unit element 1.

To close this introduction, we briefly outline the contents of
the paper. In Section 2, we begin by recalling the necessary
background  on graded Poisson color algebras.
Section 3  develops the technique of connections of elements in the support of the
grading for graed Poisson color algebras. We also show
that such an arbitrary graded Poisson color algebra $\pp$ with a  symmetric $\Lam$-support $\Sigma_{\Lam}$  is of the form
$\pp=\uu\oplus\sum_{[\lam]\in\Sigma/\sim}I_{[\lam]},,$ with $\uu$   a vector space complement  $\uu$ of $span_{\bbbf}\{ \pp_{\mu}\pp_{\mu^{-1}}+\{\pp_{\mu}, \pp_{\mu^{-1}} \} : \mu\in[\lam]\}$ in $\pp_1,$
 and any
$I_{[\lam]},$ a well-described  ideal of $\pp,$
satisfying $[I_{[\lam]}, I_{[\mu]}]+I_{[\lam]} I_{[\mu]})=0$ if $[\lam]\neq[\mu].$  In
section 4, we show that  under certain conditions, in the case of
$\pp$ being of maximal length, the gr-simplicity of the algebra is characterized.\\

\section{Preliminaries} \setcounter{equation}{0}\

Let us begin with some definitions concerning graded algebraic
structures. For a detailed discussion of this subject, we refer
the reader to the literature \cite{WGZ}.  Let $G$ be any additive
 abelian  group, a vector space $V$ is
said to be  $G$-graded, if there is a family
$\{V_g\}_{g\in G},$ vector subspaces such that
$V=\bigoplus_{g\in G}V_g.$ An element $x\in V$ is said to be homogeneous of degree $g\in G$  if $x\in V_g.$  We
denote by $\hh(V)$  the set of all homogeneous elements in $V.$   An algebra  $(\aa, \mu)$  is said
to be $G$-graded if  its underlying vector space is $G$-graded i.e. $\aa=\bigoplus_{g\in G}\aa_g$  and if furthermore $\mu(\aa_g, \aa_h)\subset\aa_{g+h}$ 
for all $g, h\in G.$

\begin{DEF}\label{bi} Let $G$ be an additive abelian group. A map $\e :
G\times G\longrightarrow\bbbk\setminus\{0\}$ is called a
 bi-character on $G$ if for all $g, h, f\in G,$
\begin{itemize}
\item[(i)] $\e(g, h)\e(h, g)=1,$

\item[(ii)] $\e(g+h, f)=\e(g, f) \e(h, f),$

\item[(iii)]  $\e(g, h+f)=\e(g, h)\e(g, f).$
\end{itemize}
\end{DEF}
The definition above implies that in particular, the following
relations hold
$$
\e(g, 0)=1=\e(0, g),~~~~\e(g, g)=1(\hbox{or}~-1),~~~\forall g\in G,
$$
where $0$ denotes the identity element of $G.$  

In this paper, for simplicity, the degree of any homogeneous element $x$ will be denoted again by $x.$ Thus if $x$ and
$y$ are two homogeneous elements of degree $g$ and $h,$ respectively and $\e$ is a bi-character on $G,$  then we shorten the notation
by writing $\e(x, y)$ instead of $\e(g, h).$
Unless stated, in the sequel all the graded spaces are over the same abelian group $G$ and the bi-character$\e$ will be the
same for all the structures.

\begin{DEF} (1) A  color algebra  is a triple $(\aa, \mu, \e)$ consisting a color vector space $\aa=\bigoplus_{g\in G}\aa_g,$ an even bilinear map  $\mu~:~  \aa\times\aa\longrightarrow\aa,$ i.e.   $\mu(\aa_g, \aa_h)\subset\aa_{g+h}$ 
for all $g, h\in G$ and a bi-character $\e$ on $G.$ We will write $\mu(x, y)=xy$ for simplicity.  The color algebra
$(\aa, \mu, \e)$  is said to be associative if $((x, y), z)=(x, (y, z))$ and $\e$-commutative if $(x, y)=\e(x, y)(y, x)$ for all $x, y, z\in\hh(\aa).$

(2) A  Lie color algebra is a color algebra $(\LL, [., .], \e)$  satisfying
\begin{itemize}
\item[(i)]$\e$-skew-symmetry, $$[x, y]=-\e(x, y)[y, x],$$

\item[(ii)] $\e$-Jacobi identity, $$[x, [y, z]]=[[x, y], z]+\e(x, y)[y, [x, z]],$$
\end{itemize}
for all $x, y, z\in\hh(\LL).$ 
\end{DEF}

\begin{DEF}\cite{WGZ} A {\em Poisson color algebra}   is a $4$-tuple  $(\PP, \mu,  \{., .\}, \e)$ consisting of a $G$-graded vector space  $\PP=\bigoplus_{g\in G}\pp_g,$   two even bilinear maps $\mu, \{., .\} : \PP\times\PP\longrightarrow\PP,$ and a  bi-character $\e :G\times G\longrightarrow\bbbk\setminus\{0\}$ satisfying the following conditions:

\begin{itemize}  
\item[(i)]  $(\PP,  \mu, \e)$ is an associative color algebra,
\item[(ii)]  $(\PP, \{., .\}, \e)$ is a   Lie color algebra,
\item[(iii)] Leibniz color identity
$$
\{xy, z\}=x\{y, z\} +\e(y, z) \{x, z\}y,
$$
 for all  $x, y, z\in\hh(\PP).$  We call the $G$-support of the grading to the set
$\Sigma_{G}~:=\{g\in G\setminus\{0\}~:~\pp_{g}\neq 0\}.$
\end{itemize}
\end{DEF}

Every Poison algebra is a  Poisson color algebra where $\e(x, y)=1$ and every  Poisson superalgebra is a Poisson color algebra in which $G=\bbbz_2$ and $\e(x, y)=(-1)^{x y}$  for any homogeneous elements $x, y\in\LL.$ Thus Poison algebras and  Poisson superalgebras are examples of Poisson color algebra.

\begin{DEF} A subalgebra $A$ of Poison color algebra $\pp$ is a $G$-graded subspace of $\pp$  such that $\{A, A\}+A A\subset A.$  A subalgebra $I$ of $\pp$ is called an ideal if $\{I, \pp\}+I \pp+\pp  I\subset I.$  We say that  $\pp$ is {\em simple} if $\{\pp,
\pp\}+\pp \pp\neq0$ and its only  (color) ideals are $\{0\}$ and $\pp.$ 
\end{DEF}

From now on,  $\pp=(\LL, \mu,  [.,.], \e)$ denotes a  Poisson  color algebra. Let us introduce the class of graded algebras in the framework of graded Poisson color algebras $\pp.$

\begin{DEF}\label{grPCA} We say that  a  Poisson  color algebra $\PP=\bigoplus_{g\in G}\pp_g,$ is a graded algebra, by means
of the abelian group $\Lam,$  if  $\pp$ decomposes as the direct sum of vector subspaces
 $\PP=\bigoplus_{\lam\in \Lam}\pp_{\lam},$ where the homogeneous spaces satisfy $\{\pp_{\lam},\PP_{\mu}\}+\pp_{\lam}\PP_{\mu}\subset\pp_{\lm\mu}$ (denoting by
juxtaposition the product in $\Lam$ with the unit element 1).  We call the $\Lam$-support of the grading to the set
$\Sigma_{\Lam}~:=\{\lam\in\Lam\setminus\{1\}~:~\pp_{\lam}\neq 0\}.$ We say that the $\Lam$-support of the grading is symmetric if $\lam\in\Sigma_{\Lam}$  implies $\lam^{-1}\in\Sigma_{\Lam}.$ 
\end{DEF}

Let  $\PP=\bigoplus_{g\in G}\pp_g$ be a $\Lam$-graded Poisson color algebra with corresponding decomposition $\PP=\pp_1\oplus(\bigoplus_{\lam\in \Sigma_{\Lam}}\pp_{\lam}).$ If we denote by $\pp_{\lam, g}~:=\pp_{\lam}\cap\pp_g$ for any $g\in G,$ then we can assert that
\begin{equation}\label{118.}
\pp=\pp_{1, 0}\oplus(\bigoplus_{g\in\Sigma_G}\bigoplus_{\lam\in \Sigma_{\Lam}}\pp_{\lam, g}),~~\hbox{with~}~\{\pp_{\lam, g}, \pp_{\mu, h}\}+\pp_{\lam, g}\pp_{\mu, h}\subset\pp_{\lam\mu, g+h},
\end{equation}
for all $\lam, \mu\in\Sigma_{\Lam}$ and all $g, h\in\Sigma_G.$

The usual regularity concepts will be understood in the graded sense. For instance, a graded ideal $I$  of $\pp$ is a color ideal which splits as $I=\bigoplus_{\lam\in \Lam} I_{\lam},$  with $I_{\lam}=I\cap\pp_{\lam}.$ The graded Poison color
algebra $\PP$ will be called gr-simple if $\{\pp, \pp\}+\pp \pp\neq0$ and its only  graded ideals are $\{0\}$ and $\pp.$

 We finally note that  split non-commutative Poisson algebra is
an example of graded Poison color algebras and so the present paper extends the results
in \cite{C0}.\\

\section{Connections in $\Sigma_\Lam$ and decompositions} \setcounter{equation}{0}\
In the following,  $\pp$ denotes a graded Poisson  color algebra with a symmetric $\Lam$-support $\Sigma_{\Lam}$ and
$\PP=\pp_1\oplus(\bigoplus_{\lam\in \Sigma_{\Lam}}\pp_{\lam}),$ the corresponding
decomposition. We begin by developing connection techniques in this framework.  

\begin{DEF}\label{conn}
Let $\lam, ~\mu$ be two elements in $\Sigma_{\Lam}.$ We say that {\em $\lam$
is $\Sigma_{\Lam}$-connected to $\mu$} if there exists a
family
$
\{\lam_1, \lam_2, \lam_3, ..., \lam_k\}\subset\Sigma_{\Lam},
$
satisfying the following conditions;\\

\begin{itemize}
\item[(1)] $\lam_1=\lam,$
\item[(2)] $\{\lam_1, \lam_1\lam_2, \lam_1\lam_2\lam_3, ..., \lam_1\lam_2\lam_3...\lam_k\}\subset\Sigma_{\Lam},$
\item[(3)] $\lam_1\lam_2\lam_3...\lam_k\in\{\mu, \mu^{-1}\}.$
\end{itemize}
The family $\{\lam_1, \lam_2, \lam_3, ..., \lam_k\}$ is called a
{\em $\Sigma_{\Lam}$-connection} from $\lam$ to $\mu.$
\end{DEF}

The next result shows that the $\Sigma_{\Lam}$-connection relation is an equivalence. Its
proof is analogous to the one for graded Lie algebras given in \cite{C3}, Proposition 2.1.

\begin{pro}\label{rel}
The relation $\sim$ in $\Sigma_{\Lam}$ defined by
$$
\lambda\sim\mu~\hbox{ if and only if}~ \lambda~\hbox{ is $\Sigma_{\Lam}$-connected to}~ \mu,
$$
is an equivalence relation.
\end{pro}

By the above proposition, we can consider the equivalence relation
in $\Sigma_{\Lam}$ by the connection relation $\sim.$ So we denote
by
$$
\Sigma_{\Lam}/\sim :=\{[\lam] : \lam\in\Sigma_{\Lam}\},
$$
where $[\lam]$ denotes  the set of elements in $\Sigma_{\Lam}$ which are
connected to $\lam.$ Clearly, if $\mu\in[\lam]$ then $\mu^{-1}\in[\lam]$ and
by Proposition \ref{rel}, if $\mu\notin[\lam]$ then
$[\lam]\cap[\mu]=\emptyset.$

\begin{lem}\label{1.} If $\mu\in [\lam]$ and $\eta, \mu\eta\in\Sigma_{\Lam}$ then $\eta, \mu\eta\in[\lam].$
\end{lem}
\noindent {\bf Proof.} Consider the $\Sigma_{\Lam}$-connection $\{\mu, \eta\},$ we get $\mu\sim\mu\eta.$ Since $\mu\in [\lam],$ by Proposition \ref{rel} we get $\mu\sim\mu\eta.$ So $\mu\eta\in[\lam].$ Next, observe that $\{\mu\eta, \mu^{-1}\}$ is a $\Sigma_{\Lam}$-connectio from $\mu\eta$ to $\eta.$ Taking into account $\mu\eta\in[\lam],$ again as above we get $\eta\in[\lam].$\qed

Our next goal in this section is to associate an adequate graded ideal
$I_{[\lam]}$ of $\pp$ to any $[\lam].$ For a fixed  $\lam\in\Sigma_{\Lam},$ we define
$$
I_{[\lam], 1} :=span_{\bbbf}\{ \pp_{\mu}\pp_{\mu^{-1}}+\{\pp_{\mu}, \pp_{\mu^{-1}} \} : \mu\in[\lam]\}\subset\pp_1.
$$
That is 
\begin{equation}\label{119}
I_{[\lam], 1}=\sum_{\mu\in[\lam],~ g\in G}(\pp_{\mu, g}
\pp_{\mu^{-1}, -g}+\{\pp_{\mu, g},
\pp_{\mu^{-1}, -g}\})\subset\pp_{1, 0}.
\end{equation}

Next, we define
$$
\vv_{[\lam]}
:=\bigoplus_{\mu\in[\lam]}\pp_\mu=\bigoplus_{\mu\in[\lam]}\bigoplus_{ g\in G }\pp_{\mu, g}.
$$
Finally, we denote by $I_{[\lam]}$ the direct sum of the two graded
subspaces above, that is,
$$
I_{[\lam]} :=I_{ [\lam], 1}\oplus\vv_{[\lam]}.
$$

\begin{pro}\label{subalg} For any
$\lam\in\Sigma_{\Lam},$ the linear subspace $I_{[\lam]}$ is a (graded)
subalgebra of $\pp.$
\end{pro}
\noindent {\bf Proof.} First, we are going to check that
$I_{[\lam]}$ satisfies $\{I_{[\lam]}, I_{[\lam]}\}\subset I_{[\lam]}.$ We have
\begin{eqnarray}\label{200}
\nonumber\{I_{[\lam]}, I_{[\lam]}\}&=&\{I_{ [\lam], 1}\oplus\vv_{[\lam]},
I_{ [\lam], 1}\oplus\vv_{[\lam]}\}\\
&\subset&\{I_{ [\lam], 1}, I_{ [\lam], 1}\}+\{I_{ [\lam], 1}, \vv_{[\lam]}\}+\{\vv_{[\lam]}, I_{ [\lam], 1}\}+\{\vv_{[\lam]}, \vv_{[\lam]}\}.
\end{eqnarray}
Let us consider the second summand in (\ref{200}). Taking into account  $  I_{ [\lam], 1}\subset\pp_1$ and $\{\pp_1, \pp_{\mu}\}\subset\pp_{\mu}$ for any $\mu\in\Sigma_{\Lam},$ we get $\{I_{ [\lam], 1}, \vv_{[\lam]}\}\subset\vv_{[\lam]},$ and by $\e$-skew-symmetry, we get 
\begin{equation}\label{201}
\{I_{ [\lam], 1}, \vv_{[\lam]}\}+\{\vv_{[\lam]}, I_{ [\lam], 1}\}\subset I_{[\lam]}.
\end{equation}

Consider now the last  summand in (\ref{200}). Suppose there exist $\mu, \eta\in[\lam]$ such that $\{\pp_{\mu}, \pp_{\eta}\}\neq 0.$ If $\mu=\eta^{-1},$  clearly
$
\{\pp_\mu, \pp_\eta\}=\{\pp_\mu, \pp_{\mu^{-1}}\}\subset I_{ [\lam], 1}.
$
Otherwise, if  $\mu\neq\eta^{-1},$ then $\mu\eta\in\Sigma_{\Lam},$ and  by Lemma \ref{1.}, one gets
$\mu\eta\in [\lam].$ Hence $\{\pp_\mu, \pp_\eta\}\subset\pp_{\mu\eta}\subset\vv_{[\lam]}.$ In any case, we have

\begin{equation}\label{203}
[\vv_{[\lam]}, \vv_{[\lam]}]\subset I_{[\lam]}.
\end{equation}
Now, consider the first summand in (\ref{200}). we have
\begin{eqnarray}\label{204}
\nonumber\{I_{ [\lam], 1}, I_{ [\lam], 1}\}&=&\bigg\{\sum_{\mu\in [\lam]}\big(\{ \pp_{\mu}\pp_{\mu^{-1}}+\{\pp_{\mu}, \pp_{\mu^{-1}} \}\big), \sum_{\eta\in [\lam]}\big(\{ \pp_{\eta}\pp_{\eta^{-1}}+\{\pp_{\eta}, \pp_{\eta^{-1}} \}\big)\bigg\}\\
&\subset&\sum_{\mu, \eta\in [\lam]}\bigg(\big\{ \pp_{\mu}\pp_{\mu^{-1}}, \pp_{\eta}\pp_{\eta^{-1}}\big\}+\big\{ \pp_{\mu}\pp_{\mu^{-1}}, \{\pp_{\eta}, \pp_{\eta^{-1}}\}\big\}\\
\nonumber&+&\big\{ \{\pp_{\mu}, \pp_{\mu^{-1}}\}, \pp_{\eta}\pp_{\eta^{-1}}\big\}+\big\{\{\pp_{\mu}, \pp_{\mu^{-1}} \}, \{\pp_{\eta}, \pp_{\eta^{-1}} \}\big\}\bigg).
\end{eqnarray}
For the first item in (\ref{204}), taking into account $\pp_{\eta}\pp_{\eta^{-1}}\subset\pp_1$ and the  Leibniz color identity,
\begin{eqnarray}\label{204.1}
\nonumber\sum_{\mu, \eta\in [\lam]}\{ \pp_{\mu}\pp_{\mu^{-1}}, \pp_{\eta}\pp_{\eta^{-1}}\}&\subset&\sum_{\mu\in [\lam]}\{ \pp_{\mu}\pp_{\mu^{-1}}, \pp_1\}\\
\nonumber&\subset&\sum_{\mu\in [\lam]}\big(\pp_\mu\{\pp_{\mu^{-1}}, \pp_1\}+\{\pp_\mu, \pp_1\}\pp_{\mu^{-1}}\big) \\
&\subset&\sum_{\mu\in [\lam]} \pp_{\mu}\pp_{\mu^{-1}}\\
\nonumber&\subset&I_{ [\lam], 1}.
\end{eqnarray}
By the $\e$-skew-symmetry, for the second and  third items in (\ref{204}), thaking into account $\pp_{\mu}\pp_{\mu^{-1}}\subset\pp_1$ and  the $\e$-Jacobi identity, 
\begin{eqnarray}\label{204.2}
\nonumber\sum_{\mu, \eta\in [\lam]}\big\{ \pp_{\mu}\pp_{\mu^{-1}}, \{\pp_{\eta}, \pp_{\eta^{-1}}\}\big\}&\subset&\sum_{\eta\in [\lam]}\big\{ \pp_1, \{\pp_{\eta}, \pp_{\eta^{-1}}\}\big\}\\
\nonumber&\subset&\sum_{\eta\in [\lam]}\bigg(\big\{\{\pp_1, \pp_\eta\}, \pp_{\eta^{-1}}\big\}+\big\{\pp_\eta, \{\pp_1, \pp_{\eta^{-1}}\}\big\}\bigg) \\
&\subset&\sum_{\eta\in [\lam]}\{ \pp_\eta, \pp_{\eta^{-1}}\}\\
\nonumber&\subset& I_{ [\lam], 1}.
\end{eqnarray}
For the last item in (\ref{204}), the $\e$-Jacobi identity,
\begin{eqnarray}\label{204.3}
\nonumber\sum_{\mu, \eta\in [\lam]}\big\{\{\pp_{\mu}, \pp_{\mu^{-1}} \}, \{\pp_{\eta}, \pp_{\eta^{-1}} \}\big\}&\subset&\sum_{\mu, \eta\in [\lam]}\bigg(\big\{ \pp_\mu, \{\pp_{\eta}, \pp_{\eta^{-1}}\}, \pp_{\mu^{-1}}\big\}\\
\nonumber&+&\big\{\pp_\mu, \{\pp_{\mu^{-1}}, \{\pp_\eta, \pp_{\eta^{-1}}\}\}\big\}\bigg)\\
\nonumber&\subset&\sum_{\eta\in [\lam]}\bigg(\big\{\{\pp_1, \pp_\eta\}, \pp_{\eta^{-1}}\big\}+\big\{\pp_\eta, \{\pp_1, \pp_{\eta^{-1}}\}\big\}\bigg) \\
&\subset&\sum_{\eta\in [\lam]}\{ \pp_\eta, \pp_{\eta^{-1}}\}\\
\nonumber&\subset& I_{ [\lam], 1}.
\end{eqnarray}
From Eqs. (\ref{204.1})-(\ref{204.3}),   we conclude that

\begin{equation}\label{205}
\{I_{[\lam], 1}, I_{[\lam], 1}\}\subset I_{[\lam], 1}.
\end{equation} 
Thus,  Eqs (\ref{201}), (\ref{203}) and (\ref{205}) give us 
\begin{equation}\label{2061}
\{I_{[\lam]}, I_{[\lam]}\}\subset I_{[\lam]}.
\end{equation} 
Next, we show that $I_{[\lam]}$ satisfies $I_{[\lam]} I_{[\lam]}\subset I_{[\lam]}.$ We have
\begin{eqnarray}\label{2071}
\nonumber I_{[\lam]} I_{[\lam]}&=&(I_{ [\lam], 1}\oplus\vv_{[\lam]})
(I_{ [\lam], 1}\oplus\vv_{[\lam]})\\
&\subset& I_{ [\lam], 1} I_{ [\lam], 1}+I_{ [\lam], 1} \vv_{[\lam]}+\vv_{[\lam]} I_{ [\lam], 1}+\vv_{[\lam]} \vv_{[\lam]}.
\end{eqnarray}
It is enough we just have to consider the first summand in (\ref{2071}). For the rest of summands, by a similar way as above, one  can shows
\begin{equation}\label{2072}
I_{ [\lam], 1} \vv_{[\lam]}+\vv_{[\lam]} I_{ [\lam], 1}+\vv_{[\lam]} \vv_{[\lam]}\subset I_{[\lam]}.
\end{equation}
Now, consider the first summand, we have
\begin{eqnarray}\label{2073}
\nonumber I_{ [\lam], 1} I_{ [\lam], 1}&=&\bigg(\sum_{\mu\in [\lam]}\big( \pp_{\mu}\pp_{\mu^{-1}}+\{\pp_{\mu}, \pp_{\mu^{-1}}\}\big)\bigg )\bigg( \sum_{\eta\in [\lam]}\big( \pp_{\eta}\pp_{\eta^{-1}}+\{\pp_{\eta}, \pp_{\eta^{-1}} \}\big)\bigg)\\
&\subset&\sum_{\mu, \eta\in [\lam]}\bigg(\big(\pp_{\mu}\pp_{\mu^{-1}}\big)\big(\pp_{\eta}\pp_{\eta^{-1}}\big)+(\pp_{\mu}\pp_{\mu^{-1}} )\{\pp_{\eta}, \pp_{\eta^{-1}}\}\\
\nonumber&+&\{\pp_{\mu}, \pp_{\mu^{-1}}\}( \pp_{\eta}\pp_{\eta^{-1}})+\{\pp_{\mu}, \pp_{\mu^{-1}} \} \{\pp_{\eta}, \pp_{\eta^{-1}} \}\bigg).
\end{eqnarray}
For the first item in (\ref{2073}), By associativity,  we have
\begin{equation*}
\sum_{\mu, \eta\in [\lam]}(\pp_{\mu}\pp_{\mu^{-1}})(\pp_{\eta}\pp_{\eta^{-1}})=\sum_{\mu, \eta\in [\lam]}\pp_{\mu}\big(\pp_{\mu^{-1}}(\pp_{\eta}\pp_{\eta^{-1}})\big)=\sum_{\mu\in [\lam]}\pp_{\mu}\pp_{\mu^{-1}}\subset I_{ [\lam], 1}.
\end{equation*}
For the second  item in (\ref{2073}), By  the  Leibniz color identity, given $\mu, \eta\in [\lam],$ we have
\begin{eqnarray*}
(\pp_{\mu}\pp_{\mu^{-1}}) \{\pp_{\eta}, \pp_{\eta^{-1}}\}&\subset&\bigg\{\big(\pp_{\mu}\pp_{\mu^{-1}}\big) \pp_{\eta}, \pp_{\eta^{-1}}\bigg\}+\bigg\{\big(\pp_{\mu}\pp_{\mu^{-1}}\big),  \pp_{\eta}\bigg\} \pp_{\eta^{-1}}\\
&\subset&\{\pp_{\eta}, \pp_{\eta^{-1}} \}+ \pp_{\eta}\pp_{\eta^{-1}}.
\end{eqnarray*}
Hence
\begin{equation*}
\sum_{\mu, \eta\in [\lam]}(\pp_{\mu}\pp_{\mu^{-1}}) \{\pp_{\eta}, \pp_{\eta^{-1}}\}\subset \sum_{\eta\in [\lam]}\big(\{\pp_{\eta}, \pp_{\eta^{-1}} \}+ \pp_{\eta}\pp_{\eta^{-1}}\big)\subset I_{ [\lam], 1},
\end{equation*}
and similarly 
\begin{equation*}
\sum_{\mu, \eta\in [\lam]}\{\pp_{\mu}, \pp_{\mu^{-1}}\}( \pp_{\eta}\pp_{\eta^{-1}})\subset I_{ [\lam], 1}.
\end{equation*}
For the last item  in (\ref{2073}), By  the  Leibniz color identity, given $\mu, \eta\in [\lam],$ we have
\begin{eqnarray*}
\{\pp_{\mu}, \pp_{\mu^{-1}} \} \{\pp_{\eta}, \pp_{\eta^{-1}} \}&\subset&\bigg\{\big\{\pp_{\mu}, \pp_{\mu^{-1}}\big\}\pp_{\eta}, \pp_{\eta^{-1}}\bigg\}+\bigg\{\big\{\pp_{\mu}, \pp_{\mu^{-1}}\big\}, \pp_{\eta}\bigg\} \pp_{\eta^{-1}}\\
&\subset&\{\pp_{\mu}, \pp_{\mu^{-1}} \}+\pp_{\mu}, \pp_{\mu^{-1}}.
\end{eqnarray*}
Hence
\begin{equation*}
\sum_{\mu, \eta\in [\lam]}\{\pp_{\mu}, \pp_{\mu^{-1}} \} \{\pp_{\eta}, \pp_{\eta^{-1}} \}\subset \sum_{\mu\in [\lam]}\big(\{\pp_{\mu}, \pp_{\mu^{-1}} \}+ \pp_{\mu}\pp_{\mu^{-1}}\big)\subset I_{ [\lam], 1}.
\end{equation*}
Thus,  we showed that the first summand  in (\ref{2071}) satisfy
\begin{equation}\label{2074}
 I_{ [\lam], 1} I_{ [\lam], 1}\subset  I_{ [\lam], 1}\subset I_{[\lam]}.
\end{equation}
Therefore, Eqs. (\ref{2072}) and (\ref{2074})  give us 
\begin{equation}\label{2075}
 I_{[\lam]} I_{[\lam]}\subset I_{[\lam]}.
\end{equation}

Finaly, from Eqs. (\ref{2061}) and (\ref{2075}), we conclude that $\{I_{[\lam]}, I_{[\lam]}\}+I_{[\lam]} I_{[\lam]}\subset I_{[\lam]},$ so $I_{[\lam]}$ is a (graded) subalgebra of $\PP.$\qed\\

\begin{pro}\label{subalg1}  For any
$\lam\in\Sigma_{\Lam},$  if $\mu\notin[\lam]$ then $\{I_{[\lam]}, I_{[\mu]}\}+I_{[\lam]} I_{[\mu]}=0.$
\end{pro}
\noindent {\bf Proof.} We have
\begin{eqnarray}\label{2}
\nonumber\{I_{[\lam]}, I_{[\mu]}\}&=&\{I_{ [\lam], 1}\oplus\vv_{[\lam]},
I_{ [\mu], 1}\oplus\vv_{[\mu]}\}]\\
&\subset&\{I_{ [\lam], 1}, I_{ [\mu], 1}\}+\{I_{ [\lam], 1}, \vv_{[\mu]}\}+\{\vv_{[\lam]}, I_{ [\mu], 1}\}+\{\vv_{[\lam]}, \vv_{[\mu]}\},
\end{eqnarray}
and also
\begin{eqnarray}\label{3}
\nonumber I_{[\lam]} I_{[\mu]}&=&(I_{ [\lam], 1}\oplus\vv_{[\lam]})
(I_{ [\mu], 1}\oplus\vv_{[\mu]})\\
&\subset& I_{ [\lam], 1} I_{ [\mu], 1}+I_{ [\lam], 1} \vv_{[\mu]}+\vv_{[\lam]} I_{ [\mu], 1}+\vv_{[\lam]} \vv_{[\mu]}.
\end{eqnarray}
We consider the last sammands in (\ref{2}) and (\ref{3}). Suppose that there exist $\lam'\in [\lam]$ and $\eta'\in[\mu]$ such that $\{\pp_{\lam'}, \pp_{\mu'}\}+\pp_{\lam'} \pp_{\mu'}\neq 0.$  As necessarily $\lam'\neq\mu'^{-1},$ then $\lam', \mu'\in\Sigma_{\Lam}.$ So $\{\lam', \mu', \lam'^{-1}\}$ is a $\Sigma_\Lam$-connection from $\lam'$ to $\mu'.$  By the transitivity of the connection relation we have $\mu'\in [\lam],$ a contradiction. Hence $\{\pp_{\lam'}, \pp_{\mu'}\}+\pp_{\lam'} \pp_{\mu'}= 0.$ and so
\begin{equation}\label{4}
\{\vv_{[\lam]}, \vv_{[\mu]}\}+\vv_{[\lam]} \vv_{[\mu]}=0.
\end{equation}
Consider now the second summands in (\ref{2}) and (\ref{3}), we have
\begin{eqnarray}\label{5}
\nonumber\{I_{ [\lam], 1}, \vv_{[\mu]}\}+I_{ [\lam], 1} \vv_{[\mu]}
&=&\bigg\{\sum_{\lam'\in[\lam]}\big(\pp_{\lam'}\pp_{\lam'^{-1}}+\{\pp_{\lam'}, \pp_{\lam'^{-1}}\}\big), \bigoplus_{\mu'\in[\mu]}\pp_{\mu'}\bigg\}\\
&+&\sum_{\lam'\in[\lam]}\big(\pp_{\lam'}
\pp_{\lam'^{-1}}+\{\pp_{\lam'},
\pp_{\lam'^{-1}}\}\big)
\big(\bigoplus_{\mu'\in[\mu]}\pp_{\mu'}\big)\\
\nonumber&\subset&\sum_{\lam'\in[\lam], \mu'\in[\mu]}\bigg(\big\{\pp_{\lam'}
\pp_{\lam'^{-1}}, \pp_{\mu'}\big\}
+\big\{\{\pp_{\lam'},
\pp_{\lam'^{-1}}\}, \pp_{\mu'}\big\}\\
\nonumber&+&(\pp_{\lam'}
\pp_{\lam'^{-1}})\pp_\mu+\{\pp_{\lam'},
\pp_{\lam'^{-1}}\}\pp_{\mu'}\bigg).
\end{eqnarray}
Given $\lam'\in [\lam]$ and $\eta'\in[\mu]$ such that
\begin{equation*}
\{\pp_{\lam'}\pp_{\lam'^{-1}}, \pp_{\mu'}\}+\big\{\{\pp_{\lam'}, \pp_{\lam'^{-1}}\}, \pp_{\mu'}\big\}+(\pp_{\lam'}\pp_{\lam'^{-1}})\pp_\mu+\{\pp_{\lam'},\pp_{\lam'^{-1}}\}\pp_{\mu'}\neq 0.
\end{equation*}
The following is divided into four situations to discuss.

{\bf Case 1.} $\{\pp_{\lam'}\pp_{\lam'^{-1}}, \pp_{\mu'}\}\neq 0.$ The Leibniz color identity gives
\begin{equation*}
0\neq\{\pp_{\lam'}\pp_{\lam'^{-1}}, \pp_{\mu'}\}\subset\pp_{\lam'}\{\pp_{\lam'^{-1}}, \pp_{\mu'}\}+\{\pp_{\lam'}, \pp_{\mu'}\}\pp_{\lam'^{-1}}.
\end{equation*}
We get either $\{\pp_{\lam'^{-1}}, \pp_{\mu'}\}\neq 0$ or $\{\pp_{\lam'}, \pp_{\mu'}\}\neq 0.$  In both cases, we have a contradiction, thanks to  equation (\ref{4}). Hence, $$\{\pp_{\lam'}\pp_{\lam'^{-1}}, \pp_{\mu'}\}= 0.$$
{\bf Case 2.} $\big\{\{\pp_{\lam'}, \pp_{\lam'^{-1}}\}, \pp_{\mu'}\big\}\neq 0.$ By $\e$-skew-symmetry and $\e$-Jacobi identity, 
\begin{equation*}
0\neq\big\{\{\pp_{\lam'}, \pp_{\lam'^{-1}}\}, \pp_{\mu'}\big\}\subset\big\{\pp_{\lam'}, \{\pp_{\lam'^{-1}}, \pp_{\mu'}\}\big\}+\big\{ \pp_{\lam'^{-1}}, \{\pp_{\mu'}, \pp_{\lam'}\}\big\}.
\end{equation*}
We get either $\{\pp_{\lam'^{-1}}, \pp_{\mu'}\}\neq 0$ or $\{\pp_{\mu'}, \pp_{\lam'} \}\neq 0.$  In both cases, we have a contradiction, thanks to  equation (\ref{4}). Hence, $$\big\{\{\pp_{\lam'}, \pp_{\lam'^{-1}}\}, \pp_{\mu'}\big\}= 0.$$
{\bf Case 3.} $(\pp_{\lam'}\pp_{\lam'^{-1}})\pp_\mu\neq 0.$ By associativity, we have
$
0\neq(\pp_{\lam'}\pp_{\lam'^{-1}})\pp_\mu=\pp_{\lam'}(\pp_{\lam'^{-1}}\pp_\mu),
$
which is a contradiction (see  equation (\ref{4})). Hence,  $$(\pp_{\lam'}\pp_{\lam'^{-1}})\pp_\mu\neq 0.$$
{\bf Case 4.}  $\{\pp_{\lam'},\pp_{\lam'^{-1}}\}\pp_{\mu'}\neq 0.$ Leibniz color identity gives
\begin{equation*}
0\neq\{\pp_{\lam'},\pp_{\lam'^{-1}}\}\pp_{\mu'}\subset\{\pp_{\lam'}\pp_{\mu'},\pp_{\lam'^{-1}}\}+\pp_{\lam'}\{\pp_{\mu'}, \pp_{\lam'^{-1}}\}.
\end{equation*}
We get either $\{\pp_{\lam'}\pp_{\mu'},\pp_{\lam'^{-1}}\}\neq 0$ or $\{\pp_{\mu'}, \pp_{\lam'^{-1}}\}\neq 0.$  In both cases, we have a contradiction, thanks to  equation (\ref{4}). Hence, $$\{\pp_{\lam'},\pp_{\lam'^{-1}}\}\pp_{\mu'}= 0.$$
Therefor,
\begin{equation}\label{6}
\{I_{ [\lam], 1}, \vv_{[\mu]}\}+I_{ [\lam], 1} \vv_{[\mu]}=0.
\end{equation}
In a similar way, we get
\begin{equation}\label{7}
\{\vv_{[\lam]}, I_{ [\mu], 1}\}+\vv_{[\lam]} I_{ [\mu], 1}=0.
\end{equation}

Finally, we consider the first summands in  (\ref{2}).  We have
\begin{eqnarray*}
 \{I_{ [\lam], 1}, I_{ [\mu], 1}\}&=&\sum_{\substack{\lam'\in[\lam]\\ \mu'\in[\mu]}}\bigg\{\big(\pp_{\lam'}\pp_{\lam'^{-1}}+\{\pp_{\lam'}, \pp_{\lam'^{-1}}\}\big), \big(\pp_{\mu'}\pp_{\mu'^{-1}}+\{\pp_{\mu'}, \pp_{\mu'^{-1}}\}\big)\bigg\}\\
&\subset& \sum_{\substack{\lam'\in[\lam]\\ \mu'\in[\mu]}}\bigg(\{\pp_{\lam'}\pp_{\lam'^{-1}}, \pp_{\mu'}\pp_{\mu'^{-1}}\}+\big\{\pp_{\lam'}\pp_{\lam'^{-1}}, \{\pp_{\mu'}, \pp_{\mu'^{-1}}\}\big\}\\
&+&\big\{\{\pp_{\mu'}, \pp_{\mu'^{-1}}\}, \pp_{\lam'}\pp_{\lam'^{-1}}\big\}+\big\{\{\pp_{\lam'}, \pp_{\lam'^{-1}}\}, \{\pp_{\mu'}, \pp_{\mu'^{-1}}\}\big\}\bigg). 
\end{eqnarray*}
Given $\lam'\in [\lam]$ and $\eta'\in[\mu]$ such that
\begin{eqnarray*}
0&\neq&\{\pp_{\lam'}\pp_{\lam'^{-1}}, \pp_{\mu'}\pp_{\mu'^{-1}}\}+\big\{\pp_{\lam'}\pp_{\lam'^{-1}}, \{\pp_{\mu'}, \pp_{\mu'^{-1}}\}\big\}\\
&+&\big\{\{\pp_{\mu'}, \pp_{\mu'^{-1}}\}, \pp_{\lam'}\pp_{\lam'^{-1}}\big\}+\big\{\{\pp_{\lam'}, \pp_{\lam'^{-1}}\}, \{\pp_{\mu'}, \pp_{\mu'^{-1}}\}\big\}.
\end{eqnarray*}
As above, Leibniz color identity, $\e$-Jacobi identity and associativity identity give us 
\begin{equation*}
\{I_{ [\lam], 1}, \vv_{[\mu]}\}+I_{ [\lam], 1} \vv_{[\mu]}+\{\vv_{[\lam]}, I_{ [\mu], 1}\}+\vv_{[\lam]} I_{ [\mu], 1}\neq 0,
\end{equation*}
a contradiction  with Eqs. (\ref{6}) and (\ref{7}).  From here,
\begin{equation}\label{8}
 \{I_{ [\lam], 1}, I_{ [\mu], 1}\}=0.
\end{equation}
In a similar manner, one can get
\begin{equation}\label{9}
 I_{ [\lam], 1} I_{ [\mu], 1}=0.
\end{equation}
From Eqs. (\ref{6})- (\ref{9}) and  (\ref{4}), we conclude the result.\qed\\

\begin{thm}\label{main1} The following assertions hold
\begin{itemize}
\item[(1)] For any $\lam\in\Sigma_{\Lam},$ the graded  Poisson  color  
subalgebra
$
I_{[\lam]} =I_{[\lam], 1}\oplus\vv_{[\lam]},
$
of $\pp$ associated to $[\lam]$ is a graded  ideal of $\pp.$
\item[(2)] If $\pp$ is gr-simple, then there exists a $\Sigma_\Lam$-connection from
$\lam$ to $\mu$ for any $\lam, \mu\in\Sigma_{\Lam},$ and
\begin{eqnarray*}
~&\pp_1=\sum_{\lam\in [\Lam]}\big(\pp_{\lam}\pp_{\lam^{-1}}
+\{\pp_{\mu}, \pp_{\lam^{-1}}\}\big)~~\hbox{or},\\
~&\pp_{1, 0}=\sum_{\lam\in[\lam],~ g\in G}\big(\pp_{\lam, g}
\pp_{\lam^{-1}, -g}+\{\pp_{\lam, g},
\pp_{\lam^{-1}, -g}\}\big).
\end{eqnarray*}
\end{itemize}
\end{thm}
\noindent {\bf Proof.}

(1) Since $\{I_{[\lam]}, \pp_1\}\subset\vv_{[\lam]}$ and $\{\vv_{[\lam]}, \pp_1\}\subset\vv_{[\lam]},$ taking into account Propositions \ref{subalg}  and  \ref{subalg1}, we have
$$
\{I_{[\lam]}, \pp\}=\big\{I_{[\lam]}, \pp_1\oplus(\bigoplus_{\mu\in[\lam]}\pp_\mu)\oplus(\bigoplus_{\eta\notin[\lam]}\pp_\eta)\big\}\subset I_{[\lam]}.
$$ 

Now, we are going to show that $I_{[\lam]}\pp\subset I_{[\lam]}.$ Observe that for any  $\mu\in[\lam],$ by associativity identity  we have
\begin{equation}\label{10}
(\pp_\mu\pp_{\mu^{-1}})\pp_1=\pp_{\mu}(\pp_{\mu^{-1}}\pp_1)\subset\pp_{\mu}\pp_{\mu^{-1}}\subset I_{[\lam], 1}.
\end{equation}
By the Leibniz color identity, we also have
\begin{equation}\label{11}
\{\pp_\mu, \pp_{\mu^{-1}}\}\pp_1=\{\pp_{\mu}\pp_1, \pp_{\mu^{-1}}\}+\pp_\mu\{\pp_1, \pp_{\mu^{-1}}\}\subset\{\pp_{\mu}, \pp_{\mu^{-1}}\}+\pp_{\mu}\pp_{\mu^{-1}}\subset I_{[\lam], 1}.
\end{equation}
From Eqs. (\ref{10}) and (\ref{11}), we get $I_{[\lam], 1}\pp_1\subset I_{[\lam], 1}$ and clearly $\vv_{[\lam]}\pp_1\subset\vv_{[\lam]},$ we conclude that 
\begin{equation}\label{12}
I_{[\lam]}\pp_1\subset I_{[\lam]}.
\end{equation}
Taking into account Propositions \ref{subalg}  and  \ref{subalg1}, we get
$$
I_{[\lam]} \pp=I_{[\lam]}\big( \pp_1\oplus(\bigoplus_{\mu\in[\lam]}\pp_\mu)\oplus(\bigoplus_{\eta\notin[\lam]}\pp_\eta)\big)\subset I_{[\lam]}.
$$ 
In a similar way as above,  we get
$$
\pp I_{[\lam]} \subset I_{[\lam]}.
$$ 
Therefore, $I_{[\lam]}$ is a graded ideal of $\pp.$

(2) The gr-simplicity of $\pp$ implies $I_{[\lam]}= \pp.$  From here, it is clear that $[\lam]=\Sigma_{\Lam}$  and
\begin{eqnarray*}
~&\pp_1=\sum_{\lam\in [\Lam]}\big(\pp_{\lam}\pp_{\lam^{-1}}
+\{\pp_{\mu}, \pp_{\lam^{-1}}\}\big)~~\hbox{or},\\
~&\pp_{1, 0}=\sum_{\lam\in[\lam],~ g\in G}\big(\pp_{\lam, g}
\pp_{\lam^{-1}, -g}+\{\pp_{\lam, g},
\pp_{\lam^{-1}, -g}\}\big).
\end{eqnarray*} \qed\\

\begin{thm}\label{main2} For a vector space complement  $\uu$ of $span_{\bbbf}\big\{ \pp_{\mu}\pp_{\mu^{-1}}+\{\pp_{\mu}, \pp_{\mu^{-1}} \} : \mu\in[\lam]\big\}$ in $\pp_1,$  we have
$$
\pp=\uu\oplus\sum_{[\lam]\in\Sigma_\Lam/\sim}I_{[\lam]},
$$
where any $I_{[\lam]}$ is one of the graded ideals of $\pp$ described in Theorem \ref{main1}-(1), satisfying $\{I_{[\lam]}, I_{[\mu]}\}+I_{[\lam]} I_{[\mu]}=0,$ whenever $[\lam]\neq[\mu].$
\end{thm}
\noindent {\bf Proof.} We have $I_{[\lam]}$  well-defined and, by Theorem  \ref{main1}-(1), a graded ideal of $\pp.$  Now, by
considering a linear complement $\u$ of $span_{\bbbf}\big\{ \pp_{\mu}\pp_{\mu^{-1}}+\{\pp_{\mu}, \pp_{\mu^{-1}} \} : \mu\in[\lam]\big\}$ in $\pp_1,$  we have
$$
\pp=\pp_1\oplus(\bigoplus_{\lam\in\Sigma_\Lam}\pp_\lam)=\uu\oplus\sum_{[\lam]\in\Sigma_\Lam/\sim}I_{[\lam]}.
$$
Finally, Proposition \ref{subalg1} gives us $\{I_{[\lam]}, I_{[\mu]}\}+I_{[\lam]} I_{[\mu]}=0,$ whenever $[\lam]\neq[\mu].$\qed\\

Let us denote by $Z(\pp)$ the centre  of $\pp,$ that is,
$Z(\pp)=\{x\in\pp~~:~~\{x, \pp\}+x \pp+\pp x=0\}.$\\

\begin{cor}\label{2.17} If $Z(\pp)=0$ and  $\pp_1=\sum_{\lam\in [\Lam]}\big(\pp_{\lam}\pp_{\lam^{-1}}
+\{\pp_{\mu}, \pp_{\lam^{-1}}\}\big).$  Then $\pp$ is the direct sum of the graded ideals
given in Theorem \ref{main1}-(1),
$$
\pp=\bigoplus_{[\lam]\in\Sigma_\Lam/\sim}I_{[\lam]},
$$
with $\{I_{[\lam]}, I_{[\mu]}\}+I_{[\lam]} I_{[\mu]}=0,$ whenever $[\lam]\neq[\mu].$
\end{cor}
\noindent {\bf Proof.} From $\pp_1=\sum_{\lam\in [\Lam]}\big(\pp_{\lam}\pp_{\lam^{-1}}
+\{\pp_{\mu}, \pp_{\lam^{-1}}\}\big),$  it is clear that
$\pp=\sum_{[\lam]\in\Sigma_\Lam/\sim}I_{[\lam]}.$ For the direct character, take some $x\in I_{[\lam]}\cap\sum_{[\mu]\in\Sigma_\Lam/\sim, [\lam]\neq [\mu]}I_{[\mu]}.$ Since $x\in I_{[\lam]},$ the fact that $\{I_{[\lam]}, I_{[\mu]}\}+I_{[\lam]} I_{[\mu]}+ I_{[\mu]}I_{[\lam]} =0$ whenever $[\lam]\neq[\mu]$ gives us
\begin{equation*}
\big\{x, \sum_{[\mu]\in\Sigma_\Lam/\sim, [\lam]\neq [\mu]}I_{[\mu]}\big\}+x\big(\sum_{[\mu]\in\Sigma_\Lam/\sim, [\lam]\neq [\mu]}I_{[\mu]}\big)+\big(\sum_{[\mu]\in\Sigma_\Lam/\sim, [\lam]\neq [\mu]}I_{[\mu]}\big) x=0.
\end{equation*}
In the other hand, since $x\in\sum_{[\mu]\in\Sigma_\Lam/\sim, [\lam]\neq [\mu]}I_{[\mu]}$ and the same above fact implies that 
$$
\{x, I_{[\lam]}\}+x I_{[\lam]}+ I_{[\lam]} x=0.
$$
That is, $x\in Z(\pp)$ and so $x=0.$
Hence, $\pp=\bigoplus_{[\lam]\in\Sigma_\Lam/\sim}I_{[\lam]}$,  as  Proposition \ref{subalg1}  $\{I_{[\lam]}, I_{[\mu]}\}+I_{[\lam]} I_{[\mu]}=0,$ whenever $[\lam]\neq[\mu].$\qed\\

\section{The graded simple components} \setcounter{equation}{0}\

In this section, we will study the  simplicity  of  graded  Poisson  color algebras and interested in studing   under which conditions a graded noncommutative Poisson color algebra decomposes as the direct sum of the family of its
gr-simple ideals.

For an ideal $I$ of  graded  Poisson  color algebra $\pp,$  the grading of $I$, assert that
\begin{equation}\label{111.}
 I=\bigoplus_{g\in G}I_g=\bigoplus_{g\in G}\bigg((I_g\cap\pp_{1, 0})\oplus\big(\bigoplus_{\lam\in\Lam} (I_g\cap\pp_{\lam, g})\big)\bigg).
\end{equation}

\begin{lem}\label{4.1} Suppose
 $Z(\pp)=0$ and  $\pp_1=\sum_{\lam\in [\Lam]}(\pp_{\lam}\pp_{\lam^{-1}}
+\{\pp_{\mu}, \pp_{\lam^{-1}}\}).$  If $I$ is an ideal of $\pp$ such that $I\subset\pp_1$ then $I=\{0\}.$
\end{lem}
\noindent {\bf Proof.} Suppose there exists a nonzero ideal $I$ of $\pp$  such that $I\subset\pp_1.$ Tthe fact that $I$ is an ideal of $\pp,$ we have
$$
\{I, \bigoplus_{\lam\in\Lam}\pp_\lam\}+I (\bigoplus_{\lam\in\Lam}\pp_\lam)+(\bigoplus_{\lam\in\Lam}\pp_\lam) I\subset I\subset\pp_1,
$$
this implies,
\begin{equation}\label{112}
\{I, \bigoplus_{\lam\in\Lam}\pp_\lam\}+I (\bigoplus_{\lam\in\Lam}\pp_\lam)+(\bigoplus_{\lam\in\Lam}\pp_\lam) I\subset I\cap(\bigoplus_{\lam\in\Lam}\pp_\lam)\subset\pp_1\cap(\bigoplus_{\lam\in\Lam}\pp_\lam)=0.
\end{equation}
Since $\pp_1=\sum_{\lam\in [\Lam]}\big(\pp_{\lam}\pp_{\lam^{-1}}
+\{\pp_{\mu}, \pp_{\lam^{-1}}\}\big),$ Eq. (\ref{112}),  the Leibniz color identity and $\e$-Jacobi identity 
give us $\{I, \pp_1\}=0.$ Again  Eq. (\ref{112}),  the Leibniz color identity and associativity give us $I\pp_1+\pp_1 I=0.$ Thus, we obtain $I\subset Z(\PP)=0.$\qed\\

Let us introduce the concepts of $\Sigma_{\Lam}$-multiplicativity and maximal length in the framework of graed Poisson color algebras, in a similar way to the ones for split non-commutative Poisson algebras (see \cite{C0}). For each $g\in G,$ we denoteby $\Lam_g :=\{\lam\in\Lam : \pp_{\lam, g}\neq 0\}.$

\begin{DEF} We say that a graded Poison color algebra $\pp$  is $\Sigma_{\Lam}$-multiplicative if given $\lam\in\Lam_{g_i}$ and $\mu\in\Lam_{g_j}$ with $g_i, g_j\in G,$ such that $\lam\mu\in\Lam,$ then
$$
\{\pp_{\lam, g_i}, \pp_{\mu, g_j}\}+\pp_{\lam, g_i}\pp_{\mu, g_j}\neq 0
$$
\end{DEF}

\begin{DEF}  A graded Poison color algebra $\pp$  is of maximal length if for
any $\lam\in\Lam_g,~g\in G,$ we have $\dim \pp_{\lam, g}=1,~~\dim \pp_{\lam^{-1}, -g}=1.$
\end{DEF}
As examples of $\Sigma_{\Lam}$-multiplicative and  of maximal length graded Poison color algebra  we have the graed Lie algebras (see \cite{C3}) and split non-commutative Poisson algebras (see \cite{C0}).

Observe that if $\pp$ is of maximal length, then equation (\ref{111.}) let us assert that given any nonzero ideal $I$ of $\pp$ then
\begin{equation}\label{39}
 I=\bigoplus_{g\in G}\big((I_g\cap\pp_{1, 0})\oplus(\bigoplus_{\lam\in\Lam_g^I} \pp_{\lam, g})\big),
\end{equation}
where $\Lam_g^I :=\{\lam\in\Lam : I_g\cap\pp_{\lam, g}\neq 0\}$ for each $g\in G.$

\begin{thm}\label{main4} Let $\pp$ be a graded Poisson color algebra of maximal length
and  $\Sigma_{\Lam}$-multiplicative. Then $\pp$ is gr-simple if and only $Z(\pp)=0,~~\pp_1=\sum_{\lam\in\Lam}\big(\pp_{\lam}\pp_{\lam^{-1}}
+\{\pp_{\mu}, \pp_{\lam^{-1}}\}\big)$  and the $\Lam$-support has all of its elements
$\Sigma_{\Lam}$-connected. 
\end{thm}
\noindent {\bf Proof.} The first implication is Theorem \ref{main1}-(2).  To prove the converse, consider $I$ a nonzero ideal of $\pp.$ By
Lemma \ref{4.1} and Eq. (\ref{39}), we can write
\begin{equation}\label{41}
 I=\bigoplus_{g\in G}\big((I_g\cap\pp_{1, 0})\oplus(\bigoplus_{\lam\in\Lam_g^I} \pp_{\lam, g})\big),
\end{equation}
with $\Lam_g^I\subset\Lam_g$ for any $g\in G$ and some $\Lam_g^I\neq\emptyset.$ Hence, we may choose $\lam_0\in\Lam_g^I$  being so
\begin{equation}\label{42}
0\neq \pp_{\lam_0, g}\subset I.
\end{equation}
Now, let us take any $\mu\in\Lam$ satisfying $\mu\notin\{\lam_0, \lam_0^{-1}\}.$ The fact that $\lam_0$ and $\mu$ are $\Sigma_{\Lam}$-connected
gives us a $\Sigma_{\Lam}$-connection $\{\lam_1, \lam_2, \lam_3, ..., \lam_k\}\subset\Sigma_{\Lam},$
satisfying the following conditions;\\

\begin{itemize}
\item[(1)] $\lam_1=\lam_0,$
\item[(2)] $\{\lam_1, \lam_1\lam_2, \lam_1\lam_2\lam_3, ..., \lam_1\lam_2\lam_3...\lam_k\}\subset\Sigma_{\Lam},$
\item[(3)] $\lam_1\lam_2\lam_3...\lam_k\in\{\mu, \mu^{-1}\}.$
\end{itemize}
Consider $\lam_1, \lam_2$ and $\lam_1\lam_2.$
 Since $\lam_2\in\Sigma_{\Lam},$ there exists $g_2\in G$  such that $\pp_{\lam_2, g_2}\neq 0$  and so $\lam_2\in\Lam_{g_2}.$ From
here, we have $\lam_1\in\Lam_{g}$ and $\lam_2\in\Lam_{g_2},$
 such that $\lam_1\lam_2\in\Lam_{g+g_2}.$
The  $\Sigma_{\Lam}$-multiplicativity and maximal
length of $\pp,$  give us either $0\neq\{\pp_{\lam_1, g}, \pp_{\lam_2, g_2}\}=\pp_{\lam_1\lam_2, g+g_2}$ or $0\neq\pp_{\lam_1, g} \pp_{\lam_2, g_2}+ \pp_{\lam_2, g_2}\pp_{\lam_1, g}=\pp_{\lam_1\lam_2, g+g_2}.$ Taking into account now that $0\neq\pp_{\lam_1, g}=\pp_{\lam_0, g}\subset I$ as consequence of
Eq. (\ref{42}) we get
$$
0\neq\pp_{\lam_1\lam_2, g+g_2}\subset I.
$$
We can argue in a similar way from $\lam_1\lam_2,~\lam_3$ and $\lam_1\lam_2\lam_3$ to get
$$
0\neq\pp_{\lam_1\lam_2\lam_3, g+g_2+g_3}\subset I~~~~~\hbox{for~ some,~~} g_3\in G.
$$
Following this process with the $\Sigma_{\Lam}$-connection $\{\lam_1, \lam_2, \lam_3, ..., \lam_k\},$ we obtain that
$$
0\neq\pp_{\lam_1\lam_2\lam_3...\lam_k, h}\subset I~~~~~\hbox{for~ some,~~} h\in G,
$$
and so either $0\neq\pp_{\mu, h}\subset I$  or $0\neq\pp_{\mu^{-1}, h}\subset I$ for some $h\in G,$ and any  $\mu\in\Lam\setminus\{\lam_0, \lam_0^{-1}\}.$

 From here along with Eq. (\ref{42})  and the fact $\pp_1=\sum_{\lam\in\Lam}\big(\pp_{\lam}\pp_{\lam^{-1}}
+\{\pp_{\mu}, \pp_{\lam^{-1}}\}\big),$ let us conclude
\begin{equation}\label{43}
\pp_1\subset I
\end{equation}
Given now any  $\lam\in\Lam,$ the fact  $\lam\neq 0,$ the maximal length of $\pp$ and Eq. (\ref{43}) 
show $\{\pp_1, \pp_\lam\}=\pp_\lam\subset I$. We conclude $I=\pp$  and therefore $\pp$ is gr-simple.\qed\\

\begin{thm} Let $\pp$ be a graded Poisson color algebra of maximal length
and  $\Sigma_{\Lam}$-multiplicative, with  $Z(\pp)=0$ and satisfying $\pp_1=\sum_{\lam\in\Lam}\big(\pp_{\lam}\pp_{\lam^{-1}}
+\{\pp_{\mu}, \pp_{\lam^{-1}}\}\big).$ Then
$$
\pp=\bigoplus_{[\lam]\in\Sigma_\Lam/\sim}I_{[\lam]},
$$
where any $I_{[\lam]}$ is a gr-simple ideal having its $\Lam$-support, $\Sigma_{I_{[\lam]}}$ with all of the elements $\Sigma_{I_{[\lam]}}$-connected.
\end{thm}
\noindent {\bf Proof.} By corollary \ref{2.17}, $\pp=\bigoplus_{[\lam]\in\Sigma_\Lam/\sim}I_{[\lam]},$ is the direct sum of the ideals
\begin{eqnarray*}
I_{[\lam]} &=&I_{[\lam], 1}\oplus\vv_{[\lam]}\\
&=&span_{\bbbf}\big\{ \pp_{\mu}\pp_{\mu^{-1}}+\{\pp_{\mu}, \pp_{\mu^{-1}} \} : \mu\in[\lam]\big\}\oplus(\bigoplus_{\mu\in[\lam]}\pp_\mu),
\end{eqnarray*}
 having any $I_{[\lam]}$  its
$\Lam$-support, $\Sigma_{I_{[\lam]}}=[\lam].$ In order to apply Theorem \ref{main4} to each $I_{[\lam]},$
observe that the fact $\Sigma_{I_{[\lam]}}=[\lam],$ gives us easily that $\Sigma_{I_{[\lam]}}$  has all of its lements $\Sigma_{I_{[\lam]}}$-connected, (connected through elements in $\Sigma_{I_{[\lam]}}).$ We also
have that any of the $I_{[\lam]}$ is $\Sigma_{I_{[\lam]}}$-multiplicative as consequence of the $\Sigma_{\Lam}$-tmultiplicativity of $\pp.$ Clearly $I_{[\lam]}$  is of maximal length. Finally, $Z_{I_{[\lam]}}(I_{[\lam]})=0,$ the centre of $I_{[\lam]}$ in itself, as consequence of 
$\{I_{[\lam]}, I_{[\mu]}\}+I_{[\lam]}, I_{[\mu]}+I_{[\mu]}, I_{[\lam]}=0$ if $[\lam]\neq[\mu]$ (see Corolary \ref{2.17}) and $Z(\pp)=0.$ We can apply Theorem \ref{main4} to any $I_{[\lam]}$ so as to conclude that
$I_{[\lam]}$ is gr-simple. It is clear that the decomposition $\pp=\bigoplus_{[\lam]\in\Sigma_\Lam/\sim}I_{[\lam]},$ satisfies the assertions of
the theorem.\qed

%%%%
%%%%

\end{document}